\documentclass{aa}  
\usepackage{color}

\usepackage{graphicx}
\usepackage{txfonts}
\begin{document}

   \title{Resonant energization of particles by radio AGN}

   \author{S. M. Mahajan\inst{1}
          \and
          Z. N. Osmanov\inst{2,3}
          }

   \institute{Institute for Fusion Studies, The University of Texas at
Austin, Austin, TX 78712, USA;
              \email{mahajan@mail.utexas.edu}
        \and
             School of Physics, Free University of Tbilisi, 0183, Tbilisi, Georgia; \email{z.osmanov@freeuni.edu.ge}
        \and     
              E. Kharadze Georgian National Astrophysical Observatory, Abastumani, 0301, Georgia
             }

   \date{Received ; accepted }

 
  \abstract
   {}
   {A new mechanism of particle acceleration, based on the resonant interaction of a classical electromagnetic wave (EM) with a quantum wave (associated with a relativistic particle), is explored.}
   {In a model illustrative calculation, we study the fate of a Klein Gordon wave subjected to the intense radio frequency waves generated in the vicinity of an active galactic nuclei (AGN). In the framework of the paper we examine a quantum wave associated with a relativistic particle, and it is shown that the group velocity of the wave approaches the speed of light, implying that the particles resonantly exchange energy with EM waves, eventually leading to acceleration of particles to very high energies.}
   {For typical parameters of under accreting Eddington radio AGN, it is shown that the resonant energization could  catapult particles to extreme energies $\sim 10^{16-20}$eV.}
   {}


   \maketitle
%

\section{Introduction} 

One of the fundamental problems of modern astrophysics is to figure the origin of particle acceleration to extremely high energies. It is very helpful to note that observations in the last decade have established/confirmed a strong correlation of very high energy (VHE) protons with active galactic nuclei \citep{kim}. In this paper, we advance a possibly acceleration mechanism operating on particles comprising the magnetospheres in nearby regions of radio AGNs. The mechanism, based on resonant energy transfer from the intense electromagnetic (EM) waves to relativistic quantum particle waves (modeled here by as Klein-Gordon Waves), is an interesting manifestation of the strong affinity between two waves with a similar mathematical structure; the dominant behavior of  both the classical electromagnetic wave (the radio waves, for example) and the relativistic Klein-Gordon wave is controlled by essentially the same hyperbolic wave operator. 

The literature is full of plausible acceleration mechanisms- the so-called Fermi mechanism \citep{fermi1} and its several variants \citep{bell,catanese} might account for the origin of extremely high energy cosmic rays; this process, though, has limitations and is efficient when the particles are initially accelerated \citep{rieger}. One of the papers relying on extreme conditions in the vicinity of compact objects \citep{bz} proposes that the poloidal magnetic field close to a black hole induces electrostatic potential, which might efficiently accelerate charged particles;  the process of acceleration, however, is limited by a significant screening effect. The so-called turbulent acceleration \citep{turb1} might account for very high energies \citep{turb2}. In the framework of this mechanism the turbulent plasma is described as a combination of Alfv\'en waves and magnetosonic modes, where the particle acceleration comes from the phase locking of trajectories of particles with the mentioned waves. As a result, this mechanism might provide energies in the PeV domain \citep{turb2}.  Magneto-centrifugal mechanism of energy pumping has been considered for AGN magnetospheres in a series of papers \citep{rieger,osm1,osm2} to explain energization of particles  up to TeV energies. Taking into account generation of centrifugally driven electrostatic field, similar mechanisms have been invoked for similar class of objects \citep{zev} - the central black hole of Milky Way, for instance \citep{sgr}. 

Harnessing of the strong wave-wave resonant interaction \citep{MA-22} for energy transfer is what distinguishes this effort from the aforementioned schemes that also operate in extreme conditions in the vicinity of highly compact objects. In particular, if the phase velocity of the EM wave coincides with the phase velocity of a quantum wave associated with a particle, the latter will  be energized to extremely high energies. The resonant enhancement is particularly pronounced when the plasma frequency is much less than the frequency of the EM radiation(Sec.2).

It is, perhaps, important to emphasize that the the manifest wave- wave interaction of this paper is possible only because of the quantum nature of the particle.

The paper is organised as follows: in Sec. 2 we outline a theory of our mechanism, consider radio-load AGN applying the theory, discuss obtained results and in Sec. 3 we summarise them.

\section{Basic Theory of Resonant energization}

In this section we will briefly outline the mechanism of resonant energization developed in  \citep{MA-22}. Enough detail is given to make this paper self contained. 


The group velocity of the quantum wave associated with a relativistic particle of energy (momentum) =E (P),  
\begin{equation}
\label{vg} 
\upsilon_g = \frac{\partial E}{\partial P} = \frac{P}{\left(P^2+m^2\right)^{1/2}},
\end{equation}
approaches the speed of light ($c = 1$) when $P\gg m$, where $P = \gamma m \upsilon$ is the momentum of the particle, $\gamma$ its relativistic factor and $\upsilon$ - the velocity. Thus for extremely high values of the momentum, the particle could, resonantly, exchange energy with an EM wave that propagates in unison. The resonant energization phenomena is well illustrated in a model calculation  \citep{MA-22} in which a Klein-Gordon (KG) wave is subjected to a circularly polarized EM wave; the latter is described by the EM four potentials $A^{\mu}$ ($A^0 = A^z = 0,$),
 \begin{equation}
\label{A} 
A^x = A\cos\left(\omega t-kz\right), \; A^y = -A\cos\left(\omega t-kz\right),
\end{equation}
where $\omega$ and $k$ are, respectively, the frequency and the wavenumber of the EM wave propagating along the $z$ axis. Because, the directions perpendicular to z are ignorable, the KG/EM system obeys the Klein-Gordon equation (see \cite{ma} for details),
\begin{equation}\label{KG} 
\left(\partial^2_t-\partial_z^2+2qAK_{\perp}\cos\left(\omega t-kz\right)+ \left(K_{\perp}^2+m^2+q^2A^2\right)\right)\Psi = 0,
\end{equation}
where $q$ is the particle charge, and $K_{\perp}$ is the wave-vector's perpendicular component, which on the other hand is the measure of the (conserved) perpendicular momentum and labels (suppressed) the wave function. The explicit $t$ and $z$ dependence in Eq. (\ref{KG}) implies  that the energy E and $P_z$ will be functions of time. The resonant solutions of Eq. (\ref{KG}) will emerge when we demand $\Psi=\Psi(\omega t-kz)\equiv\Psi(\xi)$ ($\xi\equiv\omega t-kz$); the result is a simple Mathew equation
\begin{equation}
\label{mathew} 
\left(\omega^2-k^2\right)\frac{d^2\psi}{d\xi^2}+\left(\mu+\lambda\cos\xi\right) = 0,
\end{equation}
with $\mu = K_{\perp}^2+m^2+q^2A^2$ and $\lambda = 2qAK_{\perp}$. 
Notice that the mathematical operator $\partial^2_t-\partial^2_z$ in Eq. (\ref{KG}) translates to  $\omega^2-k^2$ (in Eq. (\ref{mathew}) when we impose a solution whose phase factor  ($\xi\equiv\omega t-kz$)is exactly that of the EM wave. In some sense this is the obvious mathematics underlying wave resonance. In particular, notice that  the signature of resonance is already explicit in Eq. (\ref{mathew}); the equation is singular since $\omega^2-k^2$ tends to zero for EM waves traveling in a tenuous medium. The implication, of course, is that nontrivial solutions must demand derivative of $\Psi$ to become commensurately large. Consequently, the energy and z momentum of the KG state must also become large since both are proportional to the $\xi$ derivative (as we will soon see).

The leading order  WKB solution of Eq. (\ref{mathew}) is given by \citep{MA-22} 
\begin{equation}
\label{psi} 
\Psi = \Psi_{0} \exp(-iS\xi+i\alpha\sin\xi),
\end{equation}
where 
\begin{equation}
\label{param} 
 S = \sqrt{\frac{K_{\perp}^2+m^2+q^2A^2}{\omega^2-k^2}},\quad \alpha = \frac{qAK_{\perp}}{\sqrt{\mu\left(\omega^2-k^2\right)}}
 \end{equation}

Following simple rules of quantum mechanics, the expectation value of the energy is
$$\langle E \rangle = \frac{i\langle\Psi^{\star}\frac{\partial\Psi}{\partial t} \rangle} {\langle\Psi^{\star}\Psi\rangle} = \frac{1}{2l}\int_{-l}^{l}\left(S+\alpha\cos\xi\right)dz = $$
\begin{equation}
\label{Erms} 
 = S+\alpha\omega\cos\omega t\; \frac{\sin kl}{kl};
\end{equation}
and the acceleration takes place in the range $z = [-l, +l]$. Note that, for this resonant state, the expectation value of the z momentum $\langle K_z\rangle=(k/\omega) \langle E \rangle $ .

In order to understand the accessibility of the high energy states, let us calculate the root mean square of the rate of energy gain
$$\bar{\frac{dE}{dt}} = \left[\frac{1}{2\pi}\int_{-\pi}^{\pi}d(\omega t)\left(\frac{d\langle E \rangle}{dt}\right)^2\right]^{1/2} = $$
\begin{equation}
\label{rate} 
 = \frac{\sqrt {2}\omega^2}{\left(\omega^2-k^2\right)^{1/2}}\frac{qAK_{\perp}}{\left(m^2+K_{\perp}^2+q^2A^2\right)^{1/2}}\frac{\sin kl}{kl}.
\end{equation}
From Eqs. (\ref{param}-\ref{rate}), it is obvious that both the energy $\langle E \rangle$ and the r.m.s rate of energy increase (acceleration) are resonantly enhanced 
by the factor $\omega/\sqrt{(\omega^2-k^2)}$. If we now invoke the standard relativistic dispersion relation for EM propagation in highly under dense plasmas (see for example \cite{ma} )
\begin{equation}
\label{DR} 
\omega^2-k^2 = \frac{\omega_p^2}{\sqrt{1+q^2A^2/m^2}},
\end{equation}
where $\omega_p = \sqrt{4\pi nq^2/m}$ denotes the plasma frequency and $n$ is the number density of plasma particles, we may derive
an explicit expression for the relativistic factor (associated with the high energy state) $\gamma = E/m\approx S/m$,  
\begin{equation}
\label{gama} 
\gamma = \frac{\omega}{\omega_p}\left(1+\frac{q^2A^2}{m^2}\right)^{1/4}\left(1+\frac{q^2A^2+K_{\perp}^2}{m^2}\right)^{1/2};
\end{equation}
the first two factors reflect resonant enhancement. If $\omega>>\omega_p$, very high energy states become accessible even for moderate value of $q^2A^2/m^2$. In terms of these parameters, the acceleration rate becomes

\begin{equation}
\label{rate2} 
\frac{dE}{dt} = \frac{\sqrt{2}\omega^2}{\omega_p}\frac{\sin kl}{kl}\frac{qAK_{\perp}\left(m^2+q^2A^2\right)^{1/4}}{\left(m^2+K_{\perp}^2+q^2A^2\right)^{1/2}}, 
 \end{equation}
which, in physical units, may be approximated as ($\sin(kL)/kL\simeq 1, qA> mc^2,\hbar K_\perp c$) 

\begin{equation}
\label{rate3} 
 \frac{dE}{dt} \simeq \frac{\sqrt{2}\hbar\omega^2}{\omega_p}\left(\frac{qA}{mc^2}\right)^{1/2}K_{\perp}c. 
\end{equation}
This completes our review of the theory of resonant energization. 

\section{Radio AGNs and Processes potentially limiting energy gain}

We are now ready to apply the essentials of this theory to the particular case of a plasma in the magnetospheres of a radio AGN that emits copious EM energy in the radio frequency range. Although we will, for this paper, work out the mechanism for the radio AGN, the resonant mechanism of particle energization pertains to any AGNs emitting in the entire electromagnetic spectrum- all the way to gamma rays.

To assess the effectiveness of the resonant energy transfer from the EM waves to the relativistic KG particle-waves, we must consider processes that will impede
the acceleration process (summed up in Eq. (\ref{rate3}). Perhaps, in the present context, the two most important impeding (cooling) processes will be :1) the inverse Compton (IC) scattering of the charged particles with the ambient photon field, and 2) synchrotron radiation when relatively strong magnetic fields are present. We will deal with them in the following subsections.

\subsection{Maximum Energy Gain limited by Inverse Compton (IC) Scattering }

We begin with the well-known expression for the IC cooling power \citep{rybicki}
\begin{equation}
\label{IC} 
P_{_{Comp}}\simeq \sigma_{_{T}}c\gamma^2U,
\end{equation}
where $\sigma_{_{T}}$ is the Thomson cross section, $U = L/(4\pi rc^2)$ is the energy density of EM emission, $L$ denotes the bolometric luminosity of the radio source and $r$ is the distance from the central object. As an example we consider the typical length-scale, where the generation of very high energy (VHE) particles might take place - $r\simeq 0.001 pc$.

As the particle relativistic factor (energy) increases during resonant energization, so would the energy loss due to  IC $( \sim \gamma^2)$. Further acceleration will, therefore be terminated as soon as the rate of energy loss balances the rate of energy gain. Equating $\bar {dE/dt}\simeq P_{_{IC}}$, gives us the maximum allowed Lorentz factor

$$ \gamma_{_{Comp}}^{max}\simeq\left(\frac{\bar dE}{dt}\times\frac{4\pi rc^2}{\sigma_{_{T}}L}\right)^{1/2}\simeq$$
$$\simeq 1.4\times 10^{15}\times\left(\frac{r}{0.001\; pc}\right)^{5/8}\times$$
\begin{equation}
\label{gic} 
\;\;\;\;\;\;\;\;\;\;\;\;\;\;\;\;\;\;\;\;\;\;\;\; \times \left(\frac{f}{10^9\;Hz}\right)^{3/4}\left(\frac{10^{43}\;  erg/s}{L}\right)^{3/8} ,
\end{equation}
where $f = \omega/2\pi$ is the frequency of EM radiation and we have used an expression of the Poynting flux in terms of the vector potential $S = A^2\omega^2/(4\pi c^3)$. We normalize the frequency by $10^9$ Hz (one should note that the {\bf frequency} interval of radio AGN is $10^{7-11}$ Hz \citep{AGN}). In Eq. (\ref{gic}) we have considered a spherically symmetric accretion model, leading to the following number density of particles \citep{shapiro}

\begin{figure}
  \centering {\includegraphics[width=10cm]{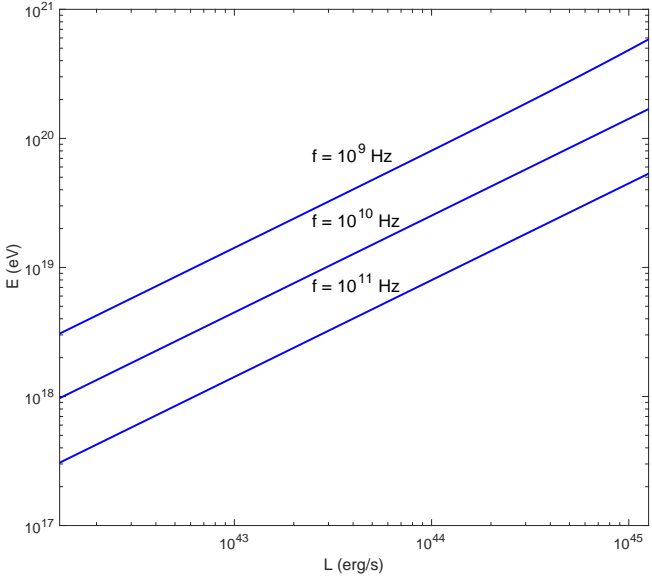}}
  \caption{Here we show the plots of the maximum energy E(L) for different emission frequencies, $f = (1; 10; 100)\times 10^9$ Hz. The  set of parameters is $M = 10^8\times M_{\odot}$, $r  =0.001$ pc, $n_{\infty}\simeq 1$ cm$^{-3}$, $T_{\infty}\simeq 10^4$ K and $\Gamma = 5/3$.}\label{fig1}
\end{figure}
\begin{figure}
  \centering {\includegraphics[width=10cm]{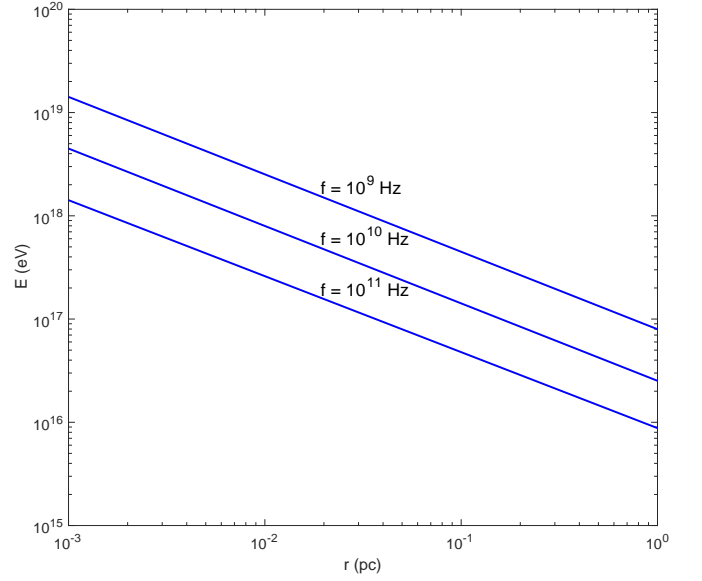}}
  \caption{{The plots of particles' energy versus distance form the central object. The  set of parameters is the same as in Fig.1, except $L = 10^{43}$ erg/s.}}\label{fig2}
\end{figure}

\begin{equation}
\label{n} 
n\simeq n_{\infty}\frac{\sqrt{2}}{4}\times\left(\frac{GM}{a_{\infty}^2r}\right)^{3/2},
\end{equation}
were $n_{\infty}\simeq 1$ cm$^{-3}$ is the number density of protons in the interstellar gas \citep{shapiro}, $G$ is the gravitational constant; the black hole is assumed to be supermassive with $M = 10^8 M_{\odot}$, where $M_{\odot}\simeq 2\times 10^{33}$ g denotes the Solar mass, and 

\begin{equation}
\label{a} 
a_{\infty} = \left(\frac{\Gamma k_BT_{\infty}}{m}\right)^{1/2}
\end{equation}
is the speed of sound of the interstellar gas and $\Gamma = 5/3$ is its adiabatic constant, $k_B$ is the Boltzmann's constant and the temperature of the interstellar gas is assumed to be $T_{\infty}\simeq 10^4$ K \citep{shapiro}.

\subsection{Maximum Energy Gain limited by synchrotron emission}

Another mechanism that potentially might limit the maximum attainable particle energy is the synchrotron radiation that is most effective when a relatively strong magnetic field is present. It is normally assumed that magnetic  energy density, $B^2/(8\pi)$, and energy density of emission, $U$ are in equipartition: $B^2/(8\pi)\simeq U$. The estimated magnetic induction, then, will be
\begin{equation}
\label{B} 
B\simeq\sqrt{\frac{2L}{r^2c}}\simeq 8.4\times\frac{0.001 pc}{r}\times \left(\frac{L}{10^{43}\;erg/s}\right)^{1/2}\; G.
\end{equation}
The upper limit on the particle energy, imposed by synchrotron cooling, will be calculated ( as in the previous case) by balancing the resonant energization rate with
\begin{equation}
\label{syn} 
P_{syn}\simeq\frac{2e^4B^2\gamma^2}{3m^2c^3},
\end{equation}
the loss rate pertaining to the relativistic regime \citep{rybicki}. We find
 $$\gamma_{syn}^{max}\simeq\left(\frac{dE}{dt}\times \frac{3m^2c^3}{2e^4B^2}\right)^{1/2}\simeq$$
$$\simeq 1.9\times10^{16}\times\left(\frac{f}{10^9\;Hz}\right)^{3/4}\times$$
\begin{equation}
\label{gsyn} 
\;\;\;\;\;\;\;\;\;\;\;\;\;\;\;\;\;\;\;\;\;\;\times\left(\frac{10^{43}\;erg/s}{L}\right)^{3/8}\left(\frac{r}{0.001 pc}\right)^{9/8} 
\end{equation}
Since $\gamma_{syn}^{max}\gg\gamma_{_{Com}}^{max}$, the synchrotron losses will not impose "any"  meaningful restrictions on maximum energy that a particle could gain via resonant energization. The relevant upper bounds might come from the Inverse Compton cooling.

On the other hand, the maximum value of the Lorentz factor, that might be provided by the present mechanism is given by Eq. (\ref{gama}), which for realistic physical quantities with $eA/mc^2>>1$ writes as
$$\gamma\simeq\frac{\omega}{\omega_p}\times\left(\frac{eA}{mc^2}\right)^{3/2}\simeq$$
\begin{equation}
\label{gama1} 
\;\;\;\;\;\;\;\simeq 8.5\times 10^7\left(\frac{10^9\;Hz}{f}\right)^{1/2}\times\left(\frac{L}{10^{43}\;erg/s}\times\frac{0.001 pc}{r}\right)^{3/4}.
\end{equation}
From this expression, after comparing to $\gamma_{_{Comp}}^{max}$ it becomes evident that inverse Compton process might significantly restrict the maximum attainable energies only for higher values of $\gamma$. Therefore, the energization process for the mentioned physical quantities is fully governed by the resonance process of acceleration.

\subsection{Conclusions-Discussion}

Equation (\ref{gama1}), giving an expression for the maximum energy constitutes the main result of this paper.  Even more, since the mass dependance of the relativistic factor is of the form $\gamma\propto 1/m$, the achieved energy  $\gamma mc^2$ does not depend on mass and therefore, the results are the same for protons and electrons.

For several representative frequencies (in the range  where the AGN spectrum peaks), the total particle energy is plotted as a function of the AGN luminosity (Fig. \ref{fig1}). The set of parameters is $f = (1; 10; 100)\times 10^9$ Hz, $M = 10^8\times M_{\odot}$, $r  =0.001$ pc, $n_{\infty}\simeq 1$ cm$^{-3}$, $T_{\infty}\simeq 10^4$ K and $\Gamma = 5/3$. From the plots it is evident that $E$ is a continuously increasing function of luminosity, which is a natural result of the relativistic factor's behaviour, $\gamma\propto L^{3/4}$, (see Eq. (\ref{gama1})).
One reads from Fig. \ref{fig1} that the newly proposed mechanism of resonant energization, can accelerate particles to several EeVs. 

In general, the number density of particles might be different for different regions of accretion matter (see Eq. (\ref{n})). In Fig.2 we plot the energy of particles versus the distance from the AGN. The luminosity is set $L =10^{43}$ erg/sec; all other parameters are the same as in Fig. 1. It is clear form the plots  that $E$, as predicted by  Eq. (\ref{gama1}): $E\propto r^{-3/4}$, is a continuously decreasing function of distance. Energies achieved by particles is of the order of $10^{16-20}$ eV, but for higher values of luminosity the achieved energies will reach  several EeVs (see Fig. 1).


As it turned out, for physical quantities considered in this manuscript the Inverse Compton process does not impose significant restrictions on particle acceleration. But 

We expect that this mechanism will operate in a variety of cosmic settings (in the vicinity of highly compact objects) and can account for some of the more energetic cosmic rays. In fact, there is a sister mechanism of resonant interaction operating directly between the KG and Gravitational waves that  could result in a similar energy transfer to fast particles \citep{am}.

\section*{Acknowledgments}
We are grateful to an anonymous referee for very interesting and valuable comments.

\section*{Data Availability}

Data are available in the article and can be accessed via a DOI link.


\end{document}